\documentclass[aps,prl,twocolumn,amsmath,amssymb,superscriptaddress]{revtex4}
\usepackage{graphicx}
\usepackage{bm}
\newcommand{\av}[1]{\langle{#1}\rangle}
\newcommand{\dr}{{\rm d}r}
\newcommand{\dx}{{\rm d}x}

\def\bfr{{\bf r}}

\def\bfR{{\bf R}}

\begin{document}
\date{\today}

\author{Heiko Gawronski}

\affiliation{Division for Atomic and Molecular Structures (ATMOS), Institute for Solid State Physics, Leibniz University of Hannover, Appelstr.\ 2, D-30167 Hannover, Germany}

\author{Jonas Fransson}
\affiliation{Department of Pysics and Astronomy, Uppsala University, Box 530, SE-7521 21 Uppsala, Sweden}

\author{Karina Morgenstern}
\affiliation{Division for Atomic and Molecular Structures (ATMOS), Institute for Solid State Physics, Leibniz University of Hannover, Appelstr.\ 2, D-30167 Hannover, Germany}

\email{Jonas.Fransson@fysik.uu.se}

\title{Real-Space Imaging of Inelastic Friedel-like Surface Oscillations Emerging from Molecular Adsorbates}

\begin{abstract}
We report real space imaging measurements of inelastic Friedel oscillations. The inelastic electron tunneling spectroscopy, using scanning tunneling microscopy, around dimers of dichlorobenze adsorbates on Au(111) surface display clear spatial modulations that we attribute to inelastic scattering at the molecular sites caused by molecular vibrations. Due to local interactions between the adsorbate and the surface states, the molecular vibrations generate a redistribution of the charge density at energies in a narrow range around the inelastic mode. Our experimental findings are supported by theoretical arguments.
\\

\end{abstract}
\maketitle


Vibrational modes are ubiquitous to molecular structures in disciplines ranging from molecular electronics and chemical analysis to biology. Inelastic scattering at these vibrational modes plays an essential role for the properties of most materials, including their electric and thermal conductivities, e.g.\ of metallic clusters \cite{hirjibehedin2006,balashov2009}, molecules \cite{stipe1998,park2000,smit2002}, superconductors \cite{liu2009}, surfaces \cite{gawronski2008,altfeder10}, carbon nanotubes and graphene \cite{gupta2006,pisana2007}.  Experimentally, inelastic transitions are probed by measuring the energy loss required to trigger the inelastic mode. Here, we present for the first time real space images of inelastic standing waves  around adsorbed molecules. These waves result from a superposition of surface state electrons before and after scattering inelastically at the adsorbate. In analogy to charge density waves arising from elastic scattering at defects (Friedel-like oscillations) \cite{friedel1958,crommie93}, we name these waves inelastic Friedel-like oscillations. Apart from the inherent beauty of such patterns, we anticipate that investigations of inelastic Friedel oscillations will provide a better understanding of vibrational excitations, coupling of electrons to molecules, and their influence on material properties.

In recent years the ability to probe quantum mechanical properties of e.g.\ single atoms \cite{hirjibehedin2006,balashov2009} and molecules \cite{stipe1998,park2000} using scanning tunneling microscopy (STM) has become remarkable. The mesurements are stretching towards atomic resolution even of collective modes as phonon excitations in metallic surfaces \cite{gawronski2008} and their transmittance from one electrode to the other one \cite{altfeder10}. These studies are of utmost importance for nano objects, for which the properties change not only from object to object, but also on the sub-object length scale. We here extend the capabilities of STM further by imaging the inelastic response of electrons scattered off nano objects.  As a model system we investigate dimers of dichlorobenzene adsorbed on Au(111) at 5 K. 

\begin{figure*}[t]
\begin{center}
\includegraphics[width=.99\textwidth]{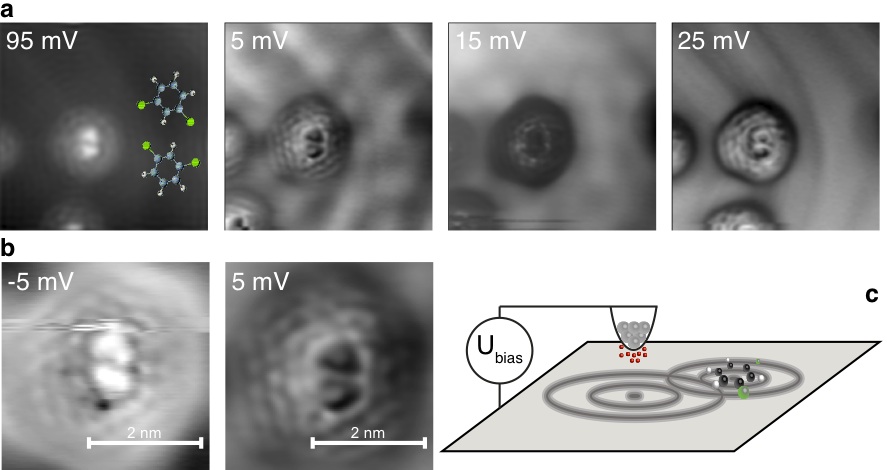}
\end{center}
\caption[]{
 {\bf a}, Leftmost panel: topography with stick-and-ball models indicating the orientation of the molecules; the Cl ligands form the two sharper edges of this triangle. Tunneling current 60 pA at 95 mV. Other panels: IETS maps recorded in constant current mode; note that the intensity scale is scaled differently on the different images in order to adjust to the different intensities of the patterns. Tunneling current 60 pA, temperature 5 K, map size: 8.7$\times$8.7 nm$^2$. {\bf b}, Zoom-in on maps at opposite polarity $\pm 5$ mV, 60 pA. {\bf c}, Schematics of the measurement principle.
}
\label{Fig1}
\end{figure*}

The dichlorobenzene molecular structure is imaged as two triangles pointing away from each other corresponding to the structure indicated by the ball-and-stick models in  \ref{Fig1}a \cite{simic08}. Inelastic electron tunneling spectroscopy (IETS) maps, shown in \ref{Fig1}a and b, represent the second derivative of the tunneling current with respect to the voltage at a certain voltage ($d^2I/dV^2$), which contains mainly the inelastic signal as detailed below. We here investigate the distinctive wave pattern emerging around the double triangle. The pattern shows three to four rings that follow the contours of the molecule pairs. In addition, these rings are modulated in intensity. The wave patterns are reminiscent of oscillations due to inelastic scattering at vibrational impurities suggested by theory \cite{fransson07}.

\begin{figure}[b]
\begin{center}
\includegraphics[width=.99\columnwidth]{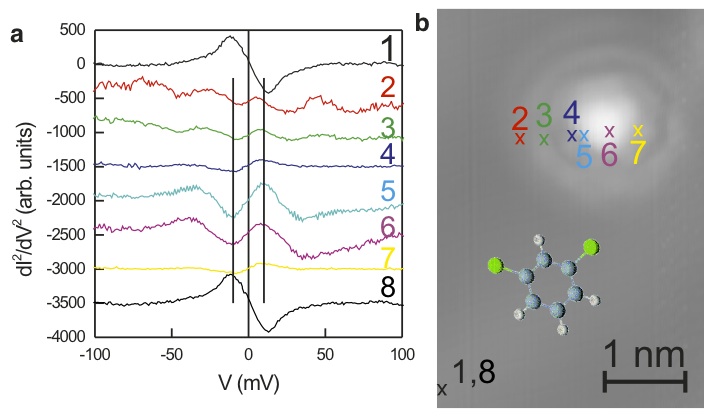}
\end{center}
\caption[]{
{\bf IETS spectra of dichlorobenze.} IETS spectra {\bf a} recorded where indicated on the STM image {\bf b}, 55 mV, 14 pA: on the surface (1,8) close to a molecule (2,3,4) and on the molecule (5,6,7) with modulation of 8 mV, 361.1 Hz;  vertical lines at $\pm 10$ mV to guide the eye. The spectra are off-set for clarity.}
\label{Fig2}
\end{figure}

The vibrational origin of these patterns is revealed in the IETS spectra on a single molecule (\ref{Fig2}). 
On the surface (spectra 1 and 8) only the characteristic phonon peak of the surface with its typical inverted symmetry (maximum  at negative and minimum at positive voltage) is observed \cite{gawronski2008}. Above the molecule (spectra 5, 6, and 7) a maximum is observed at 10 meV. The point symmetry of this peak with respect to the Fermi energy identifies it as of vibrational origin. At this low energy, external vibrations of the molecule towards the surface are expected. The broadness of the peak indicates that several modes are involved here. 
A peak in the same energy range, though with lower intensity, is observed within the region of the halo. The standing wave pattern in the IETS maps is in fact most clearly observed in the energy region of this peak. First principles density function theory calculations performed on the molecular structure reveal that there are rotation and tilting modes around 3 to 7 meV and C-Cl torsion and stretch modes between 21 and 52 meV.\cite{matsunpub} All those modes have a non-negligible coupling strength to the surface, and it is therefore likely that  modes of this kind are responsible for the patterns in our IETS maps.

We conjecture that the interference pattern is of inelastic origin. We explain it as arising from the superposition of electron wave-functions before and after scattering at the nano-object as sketched in Fig.\ \ref{Fig1}c. 
In contrast to conventional Friedel-like oscillations, however, for which the electrons are elastically scattered and thereby the incoming and outgoing electrons have the same energy, here scattering is inelastic. Therefore, the energy of the outgoing wave is reduced (or increased) by one quantum of energy corresponding to the energy of a vibrational mode.


Interpreting the standing waves in terms of inelastic scattering is theoretically corroborated by calculating a $d^2I/dV^2$ map equivalent based on the inelastic scattering assumption in a real space Green function approach. The experimental map visualizes, the inelastic signal measured in the second voltage derivative of the tunneling current, $d^2I(\bfr,V)/dV^2$, at the position $\bfr$ and voltage $V$. Theoretically, we relate the inelastic signal to the energy derivative of the local density of electron states, $\partial N(\bfr,\omega)/\partial\omega$. The local density of electron states (DOS), $N(\bfr,\omega)$, represents the quantum mechanical probability for adding or removing an electron with energy $\omega$ at a point $\bfr$ in real space. More details about the computational method are given in the Supplementary Information. Calculating the local density of states and its derivative plays the role as the theoretical counterpart to the scanning tunneling microscopy technique. 

In short, we model the inelastic signal in the measurements in terms of the real space Green function by means of coupling of the surface electrons to the local vibrational scattering centers in the spirit of Holstein coupling \cite{holstein59}. The total Green function, $G(\bfr,\bfr';\omega)$, provides the conditional probability for removing an electron with energy $\omega$ at the position $\bfr$ given that an electron with the same energy was created at $\bfr'$.  Scattering of the electron at a local vibrational mode at the position $\bfR$ by the electron is picked up by the Green function self-energy $\Sigma(\bfR,\omega)$, which provides satellite signatures in a narrow energy range around the inelastic modes $\pm\omega_0$ centered symmetrically around the Fermi level $E_F$, corresponding to the experimental observation.

\begin{figure}[t]
\begin{center}
\includegraphics[width=.99\columnwidth]{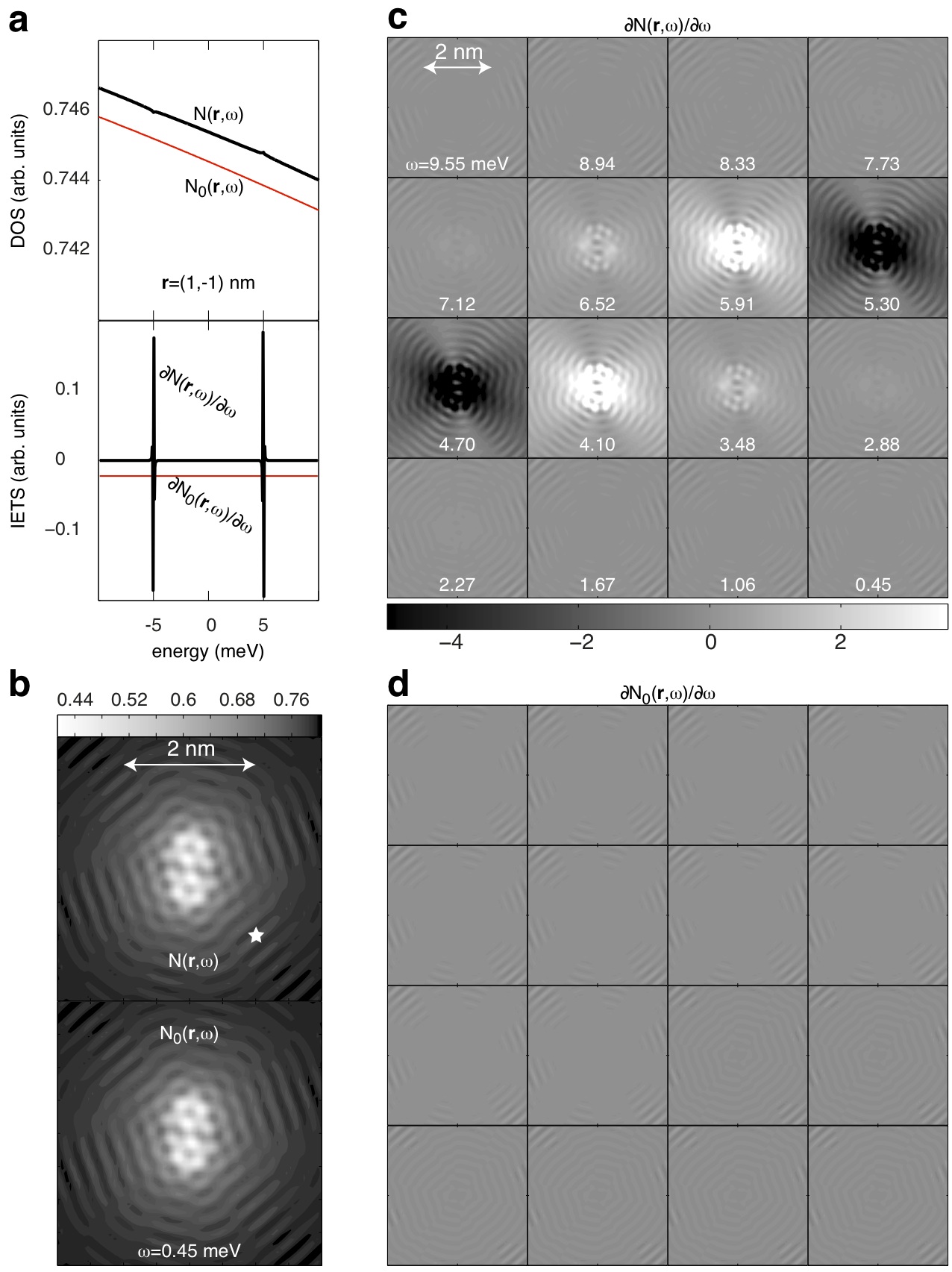}
\end{center}
\caption[]{
{\bf Conceptual comparison of the charge densities.}
{\bf a}, Energy dependence of the charge densities with and without inelastic contribution, $N(\bfr,\omega)$ and $N_0(\bfr,\omega)$ (upper panel), and their corresponding derivatives (lower panel). $N_0(\bfr,\omega)$ and its derivative have been shifted downwards for clarity.
{\bf b}, Spatial dependence of charge densities with (upper panel) and without (lower panel) inelastic contribution. The star marks the point at which the spectra plotted in panel {\bf a} were calculated.
{\bf c}, {\bf d}, IETS maps, i.e. $\partial N(\bfr,\omega)/\partial\omega$ and $\partial N_0(\bfr,\omega)/\partial\omega$ for different energies.
\label{Fig3}} 
\end{figure}

The central concept of the IETS measurement, $d^2I(\bfr,V)/dV^2$ or $\partial N(\bfr,\omega)/\partial\omega$, is based on the fact that contributions to the tunneling conductance (DOS), which are slowly varying with energy will be suppressed upon differentiation, i.e.\ in the IETS signal. On the other hand, features which are rapidly varying with energy will be enhanced by the differentiation. This is demonstrated in \ref{Fig3}a, in which we plot the calculated DOS and corresponding IETS signal for the dressed (including inelastic scattering) and bare (without inelastic scattering) system. Hence, although elastic Friedel oscillations may be recorded in the tunneling conductance (see \ref{Fig3}b, lower panel) such oscillations are suppressed in the IETS signal since the effects related to this type of scattering are slowly varying with energy (see \ref{Fig3}d). Likewise, oscillations recorded in the IETS signal have to be related to contributions in the DOS which vary on a small energy scale as illustrated by the sequence of IETS maps in \ref{Fig3}c. In particular, the inelastic scattering occurring due to a single molecular vibration provides a finite contribution to the DOS only in a narrow energy range around the inelastic mode, at which we expect the appearance of inelastic Friedel oscillations. However, due to limitation of the experimental resolution, the recorded maps display an integrated average over an energy range of 10 meV, a range in which often more than one vibrational mode is found. This means that we are presently unable to pin down the energy of each specific inelastic mode. Despite this averaging, it can be observed in \ref{Fig1}a, that the IETS maps changes with energy, which is in agreement with the theoretical arguments. Theory thus proves that the observed pattern is of inelastic origin.

\begin{figure}[t]
\begin{center}
\includegraphics[width=.99\columnwidth]{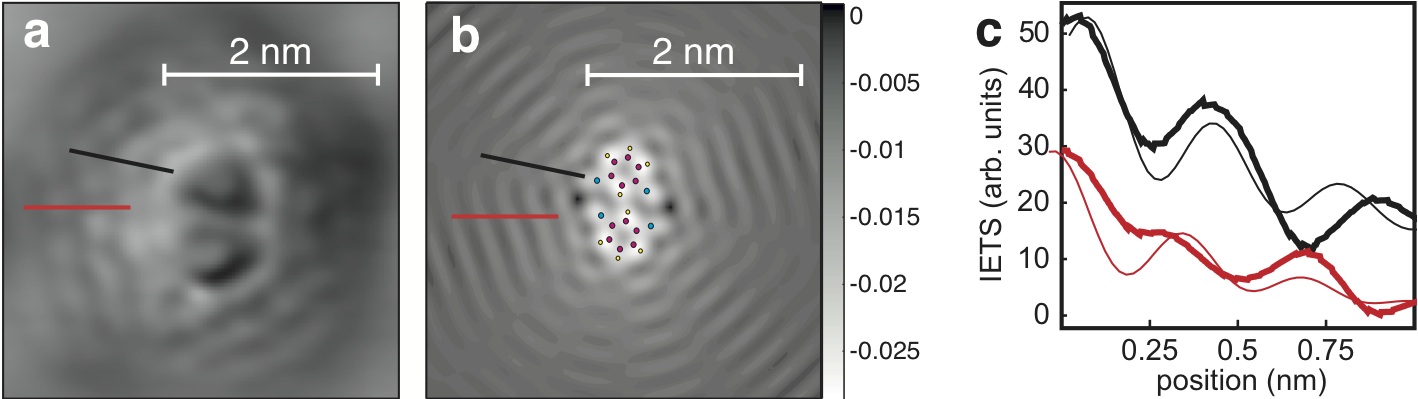}
\end{center}
\caption[]{
{\bf Comparison of calculated and experimental maps.}
{\bf a}, Experimental $d^2I/dV^2$ map at $(5 \pm 5)$ mV.
{\bf b}, Theoretical IETS map at 6 mV.
{\bf c} Line scans along {\bf c} radial directions with respect to the molecular structure as indicated in {\bf a}, {\bf b}, from theoretical (faint line) and experimental (bold line) maps. Line scans are off-set for clarity.
\label{Fig4}} 
\end{figure}

We now compare an experimental IETS image to one calculated for the specific geometry of two dichlorobenzene molecules (Fig.\ \ref{Fig4}). On the gross scale, both images show a pattern with an envolope of a truncated rhombus. On a scale closer to the molecules, there is a finer pattern of dips and peaks. When comparing line scans taken in radial directions, see \ref{Fig4}c, we find a fair agreement close to the molecular structure. The deviations at larger distances are explained by mixing with waves emerging from the surroundings. The general trend of a regular modulation with decreasing amplitude is nicely reproduced for up to four maxima. The real space decay of the waves emanating from the molecular structure is captured by $[H_0^{(1)}(k_0|\bfr-\bfR|)]^2\sim\cos(2k_0|\bfr-\bfR|)/(2k_0|\bfr-\bfR|)$ for large $k_0|\bfr-\bfR|$, where $H_0^{(1)}(x)$ is the Hankel function whereas $k_0$ is the wave vector corresponding to the inelastic mode $\omega_0$.

Combination of theory and experiment thus resolve that the inelastic Friedel-like pattern originates from the electrons sampling the molecular excitation spectrum. The signature of the molecular vibration is carried over to the electronic density of the metal and leads to standing waves around the scattering centers. These inelastic Friedel-like oscillations are spatially mapped by recording the inelastic signal, $d^2I/dV^2$. The sensitivity of the STM tip towards this signal is enhanced by functionalization via picking up a dichlorobenzene molecule (cf. \cite{gross109,gross209}).

In this paper we have demonstrated a system which fulfill the requirements for observation of inelastic Friedel-like oscillations emerging from a vibrational adsorbate on metal surface. The inelastic Friedel-like oscillations are not restricted to the specific system investigated here, which was also discussed in Ref. \cite{fransson07}. First, they should be observable on all surfaces that support an occupied surface state, i.e.\ on all surface, on which elastic Friedel-like oscillations have been reported, see e.g. Refs. \cite{crommie93,sprunger1997}. Second, all nano-objects that can be  excited vibrationally should be able to scatter the electrons inelastically. This comprises certainly a plethora of molecules, but also metallic nano-clusters, that support phonon modes. We propose that in future it will be possible to investigate the inelastic properties of such nanoobjects can be investigated by the method lined out in this article.

\section{Acknowledgements}
We thank M. Persson for communicating results prior to publication. J.F. is grateful for fruitful discussions with A. V. Balatsky, and for financial support from the Swedish Research Council (2007-562). H.G. and K.M. were supported by the Deutsche Forschungsgemeinschaft.



\end{document}